\documentclass[10pt]{iopart} 
\DeclareSymbolFont{AMSb}{U}{msb}{m}{n}
\DeclareMathSymbol{\R}{\mathalpha}{AMSb}{"52}
\begin{document}
\title[Sufficient conditions for the existence of bound states]{Sufficient conditions for the existence of bound states in a central potential}
\author{Fabian Brau\footnote[1]{E-Mail: fabian.brau@umh.ac.be}}
\address{Service de Physique G\'en\'erale et de Physique des Particules \'El\'ementaires, Groupe de Physique Nucl\'eaire Th\'eorique, Universit\'e de Mons-Hainaut, B-7000 Mons, Belgique}
\date{\today}

\begin{abstract}
We show how a large class of sufficient conditions for the existence of bound states, in non-positive central potentials, can be constructed. These sufficient conditions yield upper limits on the critical value, $g_{\rm{c}}^{(\ell)}$, of the coupling constant (strength), $g$, of the potential, $V(r)=-g v(r)$, for which a first $\ell$-wave bound state appears. These upper limits are significantly more stringent than hitherto known results. 
\end{abstract}

\maketitle

\section{Introduction}
\label{intro}

There exist in the literature several necessary conditions for the existence of at least one $\ell$-wave bound state in a given central potential. These necessary conditions yield lower limits on the critical value, $g_{\rm{c}}^{(\ell)}$, of the coupling constant (strength), $g$, of the potential, $V(r)=-g v(r)$, for which a first $\ell$-wave bound state appears.

In 1976, Glaser {\it et al.} have obtained a strong necessary condition for the existence of bound states in an arbitrary central potential in three dimensions ($\hbar^2/(2m)=1$) \cite{gla76}
\begin{equation}
\label{eq1}
\frac{(p-1)^{p-1}\,\Gamma (2p)}{(2\ell+1)^{2p-1}\,p^p\,
\Gamma^2(p)}\int_{0}^{\infty}\frac{dr}{r}\,
\left[r^2\, V^-(r)\right]^p \geq 1,
\end{equation}
where $V^-(r)=\max(0,-V(r))$ is the negative part of the potential and with the restriction $p\geq1$. This inequality is nontrivial provided that the potential $V(r)$ is less singular than the inverse square radius at the origin and that it vanishes asymptotically faster than the inverse square radius, say (for some positive $\varepsilon$)
\begin{eqnarray}
\label{eq1ba}
\lim_{r\rightarrow 0}\left[r^{2-\varepsilon}\, V(r)\right]&=&0, \\
\label{eq1bb}
\lim_{r\rightarrow \infty}\left[r^{2+\varepsilon}\, V(r)\right]&=&0.
\end{eqnarray}
We assume throughout that the potentials satisfy the relations (\ref{eq1ba}) and (\ref{eq1bb}) and that they are piecewise continuous for $r \in\ ]0,\infty[$. The lower limit on $g_{\rm{c}}^{(\ell)}$ obtained from (\ref{eq1}) is actually very accurate as it has been demonstrated on several examples (see for example \cite{gla76,las97,bra03} as well as Section \ref{sec3}).

Recently other strong necessary conditions have also been obtained \cite{bra03}
\begin{equation}
\label{eq2}
\fl \frac{2}{(2\ell+1)^2}\int_0^{\infty} dx\, x^{-2\ell}\, V^-(x)\int_0^{x} dy\, y^{2\ell+2}\, V^-(y)\geq 1,
\end{equation}
\begin{equation}
\label{eq3}
\fl \frac{6}{(2\ell+1)^3}\int_0^{\infty} dx\, x^{-2\ell}\, V^-(x)\int_0^{x}dy\, y\, V^-(y) \int_0^{y} dz\, z^{2\ell+2}\, V^-(z)\geq 1.
\end{equation}
As shown in \cite{bra03}, these two inequalities, (\ref{eq2}) and (\ref{eq3}), are natural extensions of the Bargmann-Schwinger necessary condition \cite{bar52,sch61} (first obtained by Jost and Pais \cite{jos51})
\begin{equation}
\label{eq4}
\frac{1}{2\ell+1}\int_0^{\infty} dx\, x\, V^-(x)\geq 1. 
\end{equation}
Actually the inequalities (\ref{eq4}), (\ref{eq2}) and (\ref{eq3}) are the first members of a sequence of necessary conditions which yield a monotonic sequence of lower limits on the critical value of the strength of the potential, $g_{\rm{c}}^{(\ell)}$, which converges to the exact critical strength \cite{bra03}. This remark implies that the inequality (\ref{eq3}) yields stronger restriction than the relation (\ref{eq2}). The complexity of each member of this sequence of necessary conditions becomes rapidly important and only the relation (\ref{eq2}) and (\ref{eq3}) can be easily used. It has been shown, with some test potentials, that the relation (\ref{eq3}) can be better than the relation (\ref{eq1}), especially for $\ell=0$ (see tests performed in Ref. \cite{bra03} and in Section \ref{sec3} below). 

Other necessary conditions for the existence of bound states can be found in the literature (see for example \cite{cal65,mar77} and for reviews see \cite{sim76,bra03b,bra03c}), but none, in general, yields stronger restrictions than (\ref{eq1}) and (\ref{eq3}).

Few sufficient conditions for the existence of a $\ell$-wave bound state in a central potential, yielding upper limits on $g_{\rm{c}}^{(\ell)}$, can be found in the literature. Let us mention two sufficient conditions found by Calogero in 1965 \cite{cal65b,cal65c}
\begin{equation}
\label{eq5}
\int_0^a dr\, r\, |V(r)|\, (r/a)^{2\ell+1}+\int_a^{\infty}dr\, r\, |V(r)|\,
(r/a)^{-(2\ell+1)}>2\ell+1,
\end{equation}
and
\begin{equation}
\label{eq6}
a\int_0^{\infty} dr\, |V(r)|\left[(r/a)^{2\ell}+(r/a)^{-2\ell}\,a^2|V(r)|\right]^{-1}>1.
\end{equation}
These two conditions apply provided the potential is nowhere positive, $V(r)=-|V(r)|$; in both of them $a$ is an arbitrary positive constant, and of course the most stringent conditions obtain by minimizing the left-hand sides of (\ref{eq5}) and (\ref{eq6}) over all positive values of $a$.

Few other sufficient conditions for the existence of bound states can be found in the literature (see \cite{las97,bra03,cha97}), but they are either quite complicated or less stringent than (\ref{eq5}) and (\ref{eq6}).

In this article, we obtain a strong sufficient condition for the existence of bound states yielding accurate restrictions on the critical strength $g_{\rm{c}}^{(\ell)}$ which improve significantly the restrictions provided by the relations (\ref{eq5}) and (\ref{eq6}).
 
\section{Sufficient condition and upper limit on the critical strength}
\label{sec2}

The idea used to derive the upper limit on $g_{\rm{c}}^{(\ell)}$ is to transform the standard eigenvalue problem obtained with the time independent Schr\"odinger equation, where the eigenvalues are the eigenenergies, into an eigenvalue problem where the eigenvalues are the critical coupling constants. These critical values of the strength of the potential correspond to the occurrence of an eigenstate with a vanishing energy. 

Following Schwinger \cite{sch61} (see also \cite{bi61}), we consider the zero energy Schr\"odinger equation that we write into the form of an integral equation incorporating the boundary conditions
\begin{equation}
\label{eq7}
u_{\ell}(r)=-\int_0^{\infty}dr'\, g_{\ell}(r,r')\, V(r')\, u_{\ell}(r'),
\end{equation}
where $g_{\ell}(r,r')$ is the Green's function of the kinetic energy operator and is explicitely given by
\begin{equation}
\label{eq8}
g_{\ell}(r,r')=\frac{1}{2\ell+1} r_<^{\ell+1}\, r_>^{-\ell},
\end{equation}
where $r_<=\min[r,r']$ and $r_>=\max[r,r']$. An important technical difficulty appears if the potential possesses some changes of sign (see relation (\ref{eq9}) below). This is overcome in the derivation of necessary conditions, or of upper bound on the number of bound states, by considering the negative part of the potential instead of the potential itself ($V(r)\rightarrow V^-(r)=\max(0,-V(r))$). Indeed, the potential $V^-(r)$ is more negative than $V(r)$ and thus a necessary condition for existence of a $\ell$-wave bound state in $V^-(r)$ is certainly a valid necessary condition for $V(r)$. This procedure can no longer be used to obtain sufficient conditions. For this reason we consider potentials that are nowhere positive, $V(r)=-g v(r)$, with $v(r)\geq 0$.
To obtain a symmetrical kernel we now introduce a new wave function as
\begin{equation}
\label{eq9}
\phi_{\ell}(r)=|V(r)|^{1/2}\, u_{\ell}(r).
\end{equation}
Equation (\ref{eq7}) becomes
\begin{equation}
\label{eq10}
\phi_{\ell}(r)=g\,\int_0^{\infty}dr'\, K_{\ell}(r,r')\, \phi_{\ell}(r'),
\end{equation}
where the symmetric kernel $K_{\ell}(r,r')$ is given by
\begin{equation}
\label{eq11}
K_{\ell}(r,r')=v(r)^{1/2}\, g_{\ell}(r,r')\,v(r')^{1/2}.
\end{equation}
The relation (\ref{eq10}) is thus an eigenvalue problem and, for each value of $\ell$, the smallest characteristic number is just the critical value $g_{\rm{c}}^{(\ell)}$. The other characteristic numbers correspond to the critical values of the strength for which a second, a third, ..., $\ell$-wave bound state appears. 
The kernel (\ref{eq11}) acting on the Hilbert space $L^2(\R)$ is an Hilbert-Schmidt operator for the class of potentials defined by (\ref{eq1ba}) and (\ref{eq1bb}). Thus this kernel satisfies the inequality
\begin{equation}
\int_0^{\infty} \int_0^{\infty} dx\, dy\, K_{\ell}(x,y)K_{\ell}(x,y) < \infty.
\end{equation}
Consequently the eigenvalue problem (\ref{eq10}) always possesses at least one characteristic number \cite[pp. 102-106]{tri65} (in general, this problem has an infinity of characteristic numbers). Note also that the kernel (\ref{eq11}) is the so-called Birman-Schwinger kernel \cite{sch61,bi61}.

Now we use the theorem (see for example \cite[pp. 118-119]{tri65}) which states that, for a symmetric (positive) Hilbert-Schmidt kernel, we have the variational principle
\begin{equation}
\label{eq13}
\max_{\varphi}[\int_0^{\infty} \int_0^{\infty}dx\,dy\, K_{\ell}(x,y)\, \varphi(x) \varphi(y)] =\frac{1}{g_{\rm{c}}^{(\ell)}},
\end{equation}
for $\varphi(r)$ satisfying
\begin{equation}
\label{eq12}
\int_0^{\infty}dr\, \varphi(r)^2=1.
\end{equation}
The maximal value is reached for $\varphi(x)=\phi^{\rm{c}}_{\ell}(x)$, where $\phi^{\rm{c}}_{\ell}(x)$ is the eigenfunction associated to $g_{\rm{c}}^{(\ell)}$. Consequently for an arbitrary normalized function, $f(x)$, we obtain the following upper limit on $g_{\rm{c}}^{(\ell)}$
\begin{equation}
\label{eq14}
g_{\rm{c}}^{(\ell)}\leq\left[\int_0^{\infty}\int_0^{\infty}dx\, dy\, K_{\ell}(x,y) f(x)\, f(y)\right]^{-1}.
\end{equation}

To apply the above theorem, we simply choose
\begin{equation}
\label{eq15}
f(r)=A \left[r^{2p-1}v(r)^p\right]^{1/2}, \quad p>0,
\end{equation}
where $A$ is a normalization factor. With the choice (\ref{eq15}), the upper limit (\ref{eq14}) reads
\begin{equation}
\label{eq16}
\fl g_{\rm{c}}^{(\ell)}\leq {\cal L}\,\int_0^{\infty} dx\, F(2p-1;x)\left[\int_0^{\infty} dx\, F(p;x) x^{-{\cal L}}\int_0^x dy\, F(p;y) y^{{\cal L}}\right]^{-1},
\end{equation}
with $F(q;x)=x^{q}\, v(x)^{(q+1)/2}$ and ${\cal L}=\ell+1/2$.

We do not consider other choices for the function $f(r)$ here since, as shown in Section \ref{sec3}, the relation (\ref{eq16}) is already very accurate. We just mention that another possible choice for monotonic potentials is $f(r)=A [v(r)(v(0)-v(r))^p]^{1/2}$. We have verified with an exponential potential, see (\ref{eq18}) below, that this choice yields a slight improvement.

The upper limit (\ref{eq16}) is optimal in the sense that it can be saturated, for $p=1$, by a Dirac-delta potential, $V(r)=-g\delta(r-R)$, which admits a bound state as soon as $g=R^{-1}$.

Obviously, the sufficient condition for the existence of a $\ell$-wave bound state, from which the upper limit (\ref{eq16}) on $g_{\rm{c}}^{(\ell)}$ is obtained, reads
\begin{equation}
\label{eq16b}
\fl \int_0^{\infty} dx\, \tilde{F}(p;x) x^{-{\cal L}}\int_0^x dy\, \tilde{F}(p;y) y^{{\cal L}} \left\{{\cal L}\,\int_0^{\infty} dx\, \tilde{F}(2p-1;x)\right\}^{-1}\geq 1,
\end{equation}
with $\tilde{F}(q;x)=x^{q}\, |V(x)|^{(q+1)/2}$, ${\cal L}=\ell+1/2$ and $p>0$.

\section{Tests}
\label{sec3}

\begin{table}
\protect\caption{Comparison between the exact values of the critical coupling constant 
$g_{\rm{c}}^{(\ell)}$ of the square well potential (\protect\ref{eq17}) for various values of $\ell$ and the lower limits on $g_{\rm{c}}^{(\ell)}$ obtained with the relations (\protect\ref{eq1}), (\protect\ref{eq3}) and (\protect\ref{eq4}), called respectively $g_{\rm{GGMT}}^{(\ell)}$, $g_{\rm{B}}^{(\ell)}$ and $g_{\rm{BS}}^{(\ell)}$ and the upper limits obtained with the formula, (\protect\ref{eq5}), (\protect\ref{eq6}) and (\protect\ref{eq16}), called respectively, $g_{\rm{C1}}^{(\ell)}$, $g_{\rm{C2}}^{(\ell)}$ and $g_{\rm{New}}^{(\ell)}$.}
\label{tab1}
\begin{center}
\begin{tabular}{cccccccc}
\hline\noalign{\smallskip}
$\ell$ & $g_{\rm{BS}}^{(\ell)}$ & $g_{\rm{B}}^{(\ell)}$ & $g_{\rm{GGMT}}^{(\ell)}$ & $g_{\rm{c}}^{(\ell)}$ & $g_{\rm{New}}^{(\ell)}$ & $g_{\rm{C1}}^{(\ell)}$ & $g_{\rm{C2}}^{(\ell)}$  \\
\noalign{\smallskip}\hline\noalign{\smallskip}
0 &  2   &  2.4662 &  2.3593  &  2.4674 & 2.4747 & 2.6667 & 4        \\
1 &  6   &  9.8132 &  9.1220  &  9.8696 & 9.9934 & 11.719 & 10.068   \\
2 &  10  &  19.895 &  18.454  &  20.191 & 20.604 & 25.413 & 20.895   \\
3 &  14  &  32.383 &  30.245  &  33.217 & 34.099 & 43.570 & 35.424   \\
4 &  18  &  47.064 &  44.425  &  48.831 & 50.357 & 66.089 & 53.519   \\
5 &  22  &  63.788 &  60.947  &  66.954 & 69.295 & 92.909 & 75.114   \\
\noalign{\smallskip}\hline
\end{tabular}
\end{center}
\end{table}

\begin{table}
\protect\caption{Same as for Table \protect\ref{tab1} but for the exponential potential (\protect\ref{eq18}). In the column $p$, we report the values of the variational parameter $p$ which optimize the upper limit (\protect\ref{eq16}).}
\label{tab2}
\begin{center}
\begin{tabular}{ccccccccc}
\hline\noalign{\smallskip}
$\ell$ & $g_{\rm{BS}}^{(\ell)}$ & $g_{\rm{B}}^{(\ell)}$ & $g_{\rm{GGMT}}^{(\ell)}$ & $g_{\rm{c}}^{(\ell)}$ & $g_{\rm{New}}^{(\ell)}$ & $g_{\rm{C1}}^{(\ell)}$ & $g_{\rm{C2}}^{(\ell)}$ & $p$\\
\noalign{\smallskip}\hline\noalign{\smallskip}
0 &  1   &  1.4422 &  1.4383  &  1.4458 & 1.4467 & 1.6755 & 1.5442  & 1.4686 \\
1 &  3   &  6.8546 &  7.0232  &  7.0491 & 7.0584 & 9.7188 & 7.7262  & 2.4313 \\
2 &  5   &  15.257 &  16.277  &  16.313 & 16.334 & 24.724 & 19.794  & 3.4103 \\
3 &  7   &  26.265 &  29.218  &  29.259 & 29.289 & 46.985 & 37.791  & 4.4015 \\
4 &  9   &  39.616 &  45.849  &  45.893 & 45.932 & 76.586 & 61.758  & 5.3874 \\
5 &  11  &  55.120 &  66.173  &  66.219 & 66.264 & 113.55 & 91.708  & 6.3804 \\
\noalign{\smallskip}\hline
\end{tabular}
\end{center}
\end{table}

In this Section, we propose to test the accuracy of the upper limit (\ref{eq16}) with four potentials: a square well potential, 
\begin{equation}
\label{eq17}
V(r)=-gR^{-2}\, \theta(1-r/R);
\end{equation}
an exponential potential
\begin{equation}
\label{eq18}
V(r)=-gR^{-2}\, \exp(-r/R);
\end{equation}
a Yukawa potential
\begin{equation}
\label{eq19}
V(r)=-g(rR)^{-1}\, \exp(-r/R);
\end{equation}
and the STIS (Shifted Truncated Inverse Square) potential
\begin{eqnarray}
\label{eq20}
V(r)&=&-g(R+r)^{-2}\quad {\rm for} \quad 0\leq r\leq \alpha R\nonumber \\
    &=& 0 \quad {\rm for} \quad r> \alpha R. 
\end{eqnarray}
In these potentials, the radius $R$ is arbitrary (but positive) and $\alpha$ is an arbitrary positive number.

The minimization of the upper limit (\ref{eq16}) over the positive values of $p$ can be performed analytically only for the square well potential. We find
\begin{equation}
\label{eq21}
g_{\rm{c}}^{(\ell)}\leq {\cal L}\left(\sqrt{{\cal L}+1}+1\right)^2.
\end{equation}

The comparisons between the exact value of the critical coupling constants of the potentials, $g_{\rm{c}}^{(\ell)}$, the  previously known upper and lower limits reported in Section \ref{intro} and the new upper limit (\ref{eq16}) is given in the Tables \ref{tab1}, \ref{tab2} and \ref{tab3} for various values of $\ell$ and for the potentials (\ref{eq17})-(\ref{eq19}). These comparisons show clearly that the new upper limit is very cogent as well as the lower limit (\ref{eq1}) obtained by Glaser {\it et al.} We have also performed other tests, that we do not report here, with nonmonotonic potentials and the results obtained are quite similar to those reported in these Tables.

In Table \ref{tab4}, we present the same comparison for the STIS potential but for $\ell=0$. For this potential, the critical coupling constant depends on $\alpha$. The value of $g_{\rm{c}}^{(0)}$ is obtained, for a given $\alpha$, by solving the following equation \cite{bra03b}
\begin{eqnarray}
\label{eq20a}
\lambda \ln(1+\alpha)+2 \arctan(\lambda)=2\pi, 
\end{eqnarray}
with $\lambda=\sqrt{4g_{\rm{c}}^{(0)}-1}$. For all values of $\alpha$ the results obtained with the new upper limit are again very stringent compared to previously known limits.

\begin{table}
\protect\caption{Same as for Table \protect\ref{tab1} but for the Yukawa potential (\protect\ref{eq19}). In the column $p$, we report the values of the variational parameter $p$ which optimize the upper limit (\protect\ref{eq16}).}
\label{tab3}
\begin{center}
\begin{tabular}{ccccccccc}
\hline\noalign{\smallskip}
$\ell$ & $g_{\rm{BS}}^{(\ell)}$ & $g_{\rm{B}}^{(\ell)}$ & $g_{\rm{GGMT}}^{(\ell)}$ & $g_{\rm{c}}^{(\ell)}$ & $g_{\rm{New}}^{(\ell)}$ & $g_{\rm{C1}}^{(\ell)}$ & $g_{\rm{C2}}^{(\ell)}$ & $p$ \\
\noalign{\smallskip}\hline\noalign{\smallskip}
0 &  1   &  1.6689 &  1.6643  & 1.6798 & 1.6826 & 2.0505 & 1.6810  & 1.7217 \\
1 &  3   &  8.5999 &  9.0384  & 9.0820 & 9.1039 & 13.390 & 10.706  & 3.1281 \\
2 &  5   &  19.553 &  21.839  & 21.895 & 21.937 & 35.255 & 28.374  & 4.5302 \\
3 &  7   &  33.931 &  40.074  & 40.136 & 40.194 & 67.914 & 54.819  & 5.9344 \\
4 &  9   &  51.368 &  63.744  & 63.809 & 63.880 & 111.42 & 90.071  & 7.3404 \\
5 &  11  &  71.615 &  92.850  & 92.918 & 92.998 & 165.80 & 134.14  & 8.7481 \\
\noalign{\smallskip}\hline
\end{tabular}
\end{center}
\end{table}

\begin{table}
\protect\caption{Same as for Table \protect\ref{tab1} but for the STIS potential (\protect\ref{eq20}) and $\ell=0$. In the column $p$, we report the values of the variational parameter $p$ which optimize the upper limit (\protect\ref{eq16}).}
\label{tab4}
\begin{center}
\begin{tabular}{ccccccccc}
\hline\noalign{\smallskip}
$\alpha$ & $g_{\rm{BS}}^{(0)}$ & $g_{\rm{B}}^{(0)}$ & $g_{\rm{GGMT}}^{(0)}$ & $g_{\rm{c}}^{(0)}$ & $g_{\rm{New}}^{(0)}$ & $g_{\rm{C1}}^{(0)}$ & $g_{\rm{C2}}^{(0)}$ & $p$ \\
\noalign{\smallskip}\hline\noalign{\smallskip}
0.1 &  227.22   &  282.11 &  269.84  & 282.26 & 283.12 & 306.01 & 440.67  & 1.2329 \\
0.5 &  13.864   &  17.613 &  16.842  & 17.626 & 17.683 & 19.311 & 24.664  & 1.2608 \\
1   &  5.1774   &  6.7253 &  6.4307  & 6.7319 & 6.7550 & 7.4520 & 8.6588  & 1.2889 \\
5   &  1.0434   &  1.4837 &  1.4214  & 1.4875 & 1.4939 & 1.7201 & 1.5799  & 1.4159 \\
10  &  0.67168  &  1.0066 &  0.96638 & 1.0107 & 1.0156 & 1.1998 & 1.0304  & 1.5004 \\
50  &  0.33882  &  0.58085 &  0.56233 & 0.58684 & 0.59085 & 0.74673 & 0.59855  & 1.7633 \\
\noalign{\smallskip}\hline
\end{tabular}
\end{center}
\end{table}

\section{Conclusions}
\label{sec4}

The sufficient condition (\ref{eq16b}) proposed in this article yields the upper limit (\ref{eq16}) on $g_{\rm{c}}^{(\ell)}$ which is analogous to the lower limit obtained three decades ago by Glaser {\it et al.} \cite{gla76}. The upper limit apply provided that the potential is nowhere positive and that it is less singular than the inverse square radius at the origin and that it vanishes asymptotically faster than the inverse square radius. We could use the method proposed in Ref. \cite{cha95} to consider potentials with some positive parts but the result would then be much less neat and then less interesting.

The method we use to derive the upper limit on the critical strength $g_{\rm{c}}^{(\ell)}$ is quite general and other (possibly more complicated) families of upper limits yielding (possibly) stronger restrictions on $g_{\rm{c}}^{(\ell)}$ could also be obtained. Indeed, the method is based on a variational principle for which a trial zero energy wave function is needed. There is no limitation on the accuracy of such a trial function, which imply that there is, in principle, no limitation on the accuracy of the upper limit on $g_{\rm{c}}^{(\ell)}$ derived with this procedure. In this article we have proposed in Section \ref{sec2} a compromise between accuracy and simplicity of the final formula. The accuracy of the upper limit on $g_{\rm{c}}^{(\ell)}$ was then tested in Section \ref{sec3} with some typical potentials. Clearly, the upper limit (\ref{eq16}) proposed in this article improves significantly the restriction on the possible values of $g_{\rm{c}}^{(\ell)}$ obtained with previously known upper limits.


\section*{References}
\begin{thebibliography}{99}
\bibitem{gla76} Glaser V, Grosse H, Martin A and Thirring W 1976 \textit{Studies in mathematical physics -- Essays in honor of Valentine Bargmann}, Princeton University Press, 1976, pp. 169.
\bibitem{las97} Lassaut M and Lombard R J 1997 \textit{J. Phys. A} \textbf{30} 2467.
\bibitem{bra03} Brau F 2003 \textit{J. Phys. A} \textbf{36} 9907.
\bibitem{bar52} Bargmann V 1952 \textit{Proc. Nat. Acad. Sci. U.S.A.} \textbf{38} 961.
\bibitem{sch61} Schwinger J 1961 \textit{Proc. Nat. Acad. Sci. U.S.A.} \textbf{47} 122.
\bibitem{jos51} Jost R and Pais A 1951 \textit{Phys. Rev.} \textbf{82} 840.
\bibitem{cal65} Calogero F 1965 \textit{Nuovo Cimento} \textbf{36} 199.
\bibitem{mar77} Martin A 1977 \textit{Commun. Math. Phys.} \textbf{55} 293.
\bibitem{sim76} Simon B 1976 \textit{Studies in mathematical physics - Essays in honor of Valentine Bargmann}, Princeton University Press, 1976, pp. 305-326.
\bibitem{bra03b} Brau F and Calogero F 2003 \textit{J. Math. Phys.} \textbf{44} 1554.
\bibitem{bra03c} Brau F and Calogero F 2003 \textit{J. Phys. A} \textbf{36} 12021.
\bibitem{cal65b} Calogero F 1965 \textit{J. Math. Phys.} \textbf{6} 161.
\bibitem{cal65c} Calogero F 1965 \textit{J. Math. Phys.} \textbf{6} 1105.
\bibitem{cha97} Chadan K and Kobayashi R 1997 \textit{J. Math. Phys.} \textbf{38} 4900.
\bibitem{bi61} Birman S 1961 {\it Math. Sb.} {\bf 55} 124 (1961); 1966 {\it Amer.
Math. Soc. Transl.} {\bf 53} 23.
\bibitem{tri65} Tricomi F G 1965 \textit{Integral equations}, Interscience Publishers, New-York, 1965, pp. 118-119.
\bibitem{cha95} Chadan K and Grosse H 1983 \textit{J. Phys. A} {\bf 16} 955.
\end{thebibliography}
\end{document}